\documentclass[fleqn,usenatbib,usedcolumn]{mnras}
\usepackage[british]{babel}             
\usepackage[T1]{fontenc}                
\usepackage{graphicx}                   
\usepackage[usenames,dvipsnames]{color}

\hypersetup{pdfauthor={I. Heywood},
               pdftitle={Field sources near the southern-sky calibrator PKS B1934-638: effect on spectral line observations with SKA-MID and its precursors},
               pdfkeywords={techniques: interferometric},
               bookmarksnumbered=true}

\volume{{\rm in press}}   

\usepackage{graphicx}	
\usepackage{amsmath}	
\usepackage{amssymb}	
\usepackage{multicol}        
\usepackage{bm}		
\usepackage{pdflscape}	

\newcommand{\trm}{\textrm}

\newcommand{\ltsim}{\mbox{{\raisebox{-0.4ex}{$\stackrel{<}{{\scriptstyle\sim}}$}}}}

\definecolor{mycolour}{cmyk}{1, 0., 0., 0.}

\usepackage[T1]{fontenc}
\usepackage{ae,aecompl}

\usepackage{newtxtext,newtxmath}

\title[VLA imaging of the XMM-LSS / VIDEO deep field]{VLA imaging of the XMM-LSS / VIDEO deep field at 1--2 GHz}

\author[Heywood et al.]
{\parbox{\textwidth}{
\begin{flushleft}
I.~Heywood$^{1,2,3}$\thanks{Contact author: {\tt ian.heywood@physics.ox.ac.uk}}, 
C.~L.~Hale$^{1,4}$,
M.~J.~Jarvis$^{1,5}$, 
S.~Makhathini$^{2,3}$,
J.~A.~Peters$^{1}$,
M.~L.~L.~Sebokolodi$^{6,2}$ and
O.~M.~Smirnov$^{2,3}$\\
\end{flushleft}
}
\footnotesize
\\
$^{1}$Astrophysics, Department of Physics, University of Oxford, Keble Road, Oxford, OX1 3RH, UK\\ 
$^{2}$Department of Physics and Electronics, Rhodes University, PO Box 94, Makhanda, 6140, South Africa\\
$^{3}$South African Radio Astronomy Observatory, 2 Fir Street, Black River Park, Observatory, Cape Town, 7925, South Africa\\
$^{4}$CSIRO Astronomy and Space Science, PO Box 1130, Bentley, WA 6102, Australia \\
$^{5}$Physics Department, University of the Western Cape, Private Bag X17, Bellville, 7535, South Africa\\
$^{6}$National Radio Astronomy Observatory, PO Box 0, Soccoro, NM 87801, USA\\
}
\date{Accepted 2020 June 10. Received 2020 June 09; in original form 2019 June 25}

\pubyear{2020}

\begin{document}
\label{firstpage}
\pagerange{\pageref{firstpage}--\pageref{lastpage}}
\maketitle

\begin{abstract}

Modern radio telescopes are routinely reaching depths where normal starforming galaxies are the dominant observed population. Realising the potential of radio as a tracer of star formation and black hole activity over cosmic time involves achieving such depths over representative volumes, with radio forming part of a larger multiwavelength campaign. In pursuit of this we used the Karl G. Jansky Very Large Array (VLA) to image $\sim$5 deg$^{2}$ of the VIDEO/XMM-LSS extragalactic deep field at 1--2 GHz. We achieve a median depth of 16 $\mu$Jy beam$^{-1}$ with an angular resolution of 4.5\arcsec. Comparisons with existing radio observations of XMM-LSS showcase the improved survey speed of the upgraded VLA: we cover 2.5 times the area and increase the depth by $\sim$20\% in 40\% of the time. Direction-dependent calibration and wide-field imaging were required to suppress the error patterns from off-axis sources of even modest brightness. We derive a catalogue containing 5,762 sources from the final mosaic. Sub-band imaging provides in-band spectral indices for 3,458 (60\%) sources, with the average spectrum becoming flatter than the canonical synchrotron slope below  1 mJy. Positional and flux-density accuracy of the observations, and the differential source counts are in excellent agreement with those of existing measurements. A public release of the images and catalogue accompanies this article.

\end{abstract}





\section{Introduction}

Traditionally, the focus of radio continuum surveys has been on finding and studying the radio-loud active galactic nuclei (AGN) populations over the history of the Universe, and more recently the impact that such radio-powerful objects have on their host galaxy and immediate environment. However, this focus is beginning to change as we move towards ever-deeper radio continuum surveys. This is due to the fact that at $S_{1.4\rm GHz} \ltsim 100$\,$\mu$Jy the composition of the radio source population begins to change from being AGN-dominated to being composed predominantly of star-forming galaxies (SFGs) and radio-quiet AGN \citep[e.g.][]{JarvisRawlings2004,White2015}. 

Deep radio-continuum surveys are therefore opening up a new window on what is usually considered the `normal' galaxy population. Radio observations of such star-forming galaxies are important, as they have the potential to provide a relatively unbiased view of the time-averaged star-formation rate (SFR). This is due to the fact that supernovae are co-located with regions of massive star formation, and when electrons traverse their ageing shock fronts they decelerate rapidly, producing synchrotron radiation \citep{Condon1992}.

Thus, over the past decade there have been many efforts to obtain deep radio continuum data over representative volumes of the Universe. Leading the way was the original VLA-COSMOS survey \citep{Schinnerer2007} which used the VLA at L-band ($\sim1.4$\,GHz). This has been succeeded by similar surveys spanning an order of magnitude in radio frequency, using the Giant Metrewave Radio Telescope \citep[GMRT; e.g.][]{Bondi2007,Ibar2009}, the Australia Telescope Compact Array \citep[ATCA; e.g.][]{Middelberg2008,Franzen2015} and the upgraded VLA which has recently been used to revisit the COSMOS field at 3~GHz \citep{Smolcic2017}. 

The key aims of surveys such as those listed above, as well as future approved and proposed surveys with the SKA and its precursors \citep[e.g.][]{MIGHTEE,EMU,JarvisSKA,PrandoniSeymour2015}, are to understand the link between AGN activity and the host galaxy properties, to trace the star-formation history of the Universe and the evolution in the star-formation main sequence \citep[e.g.][]{Noeske2007, Daddi2007, Whitaker2012, Johnston2015}. Radio observations are free from obscuration by dust (cf. ultraviolet tracers), and are generally not confused and/or suffer from low-angular resolution (cf. \emph{Herschel} and SCUBA-2 surveys). 

In order to achieve these goals the radio observations must target fields in the sky with excellent multi-wavelength coverage. This is because the radio data alone do not provide any information on the stellar mass of the galaxies or their redshifts, although this may soon change at least at $z<0.6$ where H{\sc i} will become a viable line for measuring redshifts from the same data as the continuum data \citep[e.g.][]{CHILES,MIGHTEE,LADUMA,Maddox2016}. 

In this paper we present a new survey with the VLA of the XMM-LSS-VIDEO field \citep{Jarvis2013} at 1-2\,GHz using the B-configuration. This survey represents one of the deepest $\sim$5 square degree surveys of the radio sky \cite[see also][]{Prandoni2018}. The target field has an exceptional range of multi-wavelength imaging at optical \citep{HSC,CFHTLS,DES}, near-infrared \citep{Jarvis2013}, mid-infrared \citep{SWIRE, SERVS}, far-infrared \citep{HerMES} and X-ray \citep{XSERVS} wavelengths, as well as spectroscopy \citep{VVDS,VUDS,DEVILS,VIPERS}, in addition to lower-frequency radio observations \citep{Smolcic2018, Hale2019a}. Our survey also significantly extends the areal coverage over this field, which also incorporates the VVDS region \citep{Bondi2003} and the UKIDSS-UDS field \citep{Simpson2006}.

The paper is organised as follows. In Section~\ref{sec:obs} we describe the observations and the data processing. In Section~\ref{sec:data} we present the data products, including a spectral index ($\alpha$)\footnote{We adopt the convention that the flux density $S$ is proportional to the observing frequency ${\nu}$ according to: $S \propto \nu^{\alpha}$.} image and a source catalogue generated using the {\sc ProFound} software. We compare these observations with previous radio surveys in Section~\ref{sec:discussion} and present the source counts derived from the survey. In Section~\ref{sec:conclusion} we briefly summarise our findings and provide a link from where the data products may be freely downloaded.

\section{Observations and data processing}
\label{sec:obs}

The observations\footnote{Project codes: 13B-308 and 15A-477, corresponding to observations taking place between 28 November -- 24 December 2013 (all pointings except 18), and on 21 April 2015 (pointing 18).} were conducted using the VLA in B-configuration. A single 1.5~h Scheduling Block (SB) was submitted for each of the 32 pointings, containing the necessary calibrator scans, as well as scans of the science target. The on-source observation time for each target pointing was 67.5~m, with 3~s per correlator integration. The correlator was configured in standard wide-band continuum mode, with 0.994--2.018 GHz of spectral coverage split up into 16~$\times$~64 MHz spectral windows (SPWs) for a total of 1,024 channels.

Data were delivered from the observatory in a {\sc CASA} format Measurement Set (MS) containing visibility data for the target and the primary and secondary calibrators. The primary calibrator was 3C147, and was used to determine the absolute flux density scale using the models derived by \citet{Perley2013} and solve for the bandpass shape. The secondary calibrator was J0217+0144 and this was used to determine time-dependent complex gain corrections. Visits to this source were somewhat infrequent, reasoning that self-calibration of the target data would be both feasible and necessary. The description of the steps that follow were applied to each SB individually. The referenced calibrator corrections were derived and applied using the NRAO {\sc CASA} pipeline\footnote{{\tt \scriptsize https://science.nrao.edu/facilities/vla/data-processing/pipeline}}. The pipeline also applied Hanning smoothing to the data, and made a first pass of automatic radio frequency interference (RFI) excision. Following this we examined the scans of the calibrator sources for any remaining RFI. Any gross features were flagged, and the pipeline was re-run from scratch. RFI was rife, with SPWs 5 and 6 (1.314--1.412 GHz) lost in many pointings, and spectral windows 8 and 9 (1.506--1.634 GHz) discarded outright for all 32 pointings. Once the reference calibration steps were complete, the corrected visibilities for the target field were split into a single source MS ready for imaging and further calibration.

All imaging was performed using using the {\sc wsclean} package \citep{Offringa2014}, which makes use of the efficient w-snapshot algorithm \citep{Humphreys2011} to correct for the effects of using non-coplanar arrays to conduct wide-field imaging. Imaging parameters were the same for each run, using 12,000~$\times$~12,000 pixels with a scale of 0.7$''$ to cover 2.33\degr~$\times$~2.33\degr. Images of this size were necessary to deconvolve and model confusing sources in the sidelobes of the primary beam. \citet{Briggs1995} weighting was used in order to suppress the sidelobes in the point spread function (PSF), with the robustness parameter set to 0.0. The frequency dependence of the sky brightness distribution was captured by deconvolving in 4~$\times$~256~MHz sub-bands. When searching for peaks of emission during the minor cycle {\sc wsclean} uses the full-band image, however deconvolution takes place in each of the sub-bands independently. A polynomial with a user-defined order (in this case 3) is fitted to the clean components and inverted into a visibility model for subtraction during the major cycle. At the end of the cleaning process the final model is (optionally) inverted and written to the {\tt MODEL\_DATA} column of the MS for use in self-calibration.

An initial imaging run was performed with unconstrained deconvolution terminating at 50,000 iterations. We then used the {\sc PyBDSF} \citep{MohanRafferty2015} source finder to locate regions of significant emission in the image. {\sc PyBDSF} works by estimating the spatial variation in the background noise level, and then identifying peaks that are some threshold (in this case 5) times the background. Once these are identified a flood fill operation takes place down to a secondary threshold (in this case 3) times the background. These islands of emission are then decomposed into groups of point and Gaussian components.

The resulting catalogue was manually examined, and spurious features were removed. Essentially all of such features were associated with residual PSF-like structures in the image which were not successfully deconvolved due to calibration deficiencies. The positions and shapes of the components in the pruned catalogue were written into a blank FITS image for subsequent use as a Boolean cleaning mask. 

Imaging was repeated with the mask being used to constrain the locations of the deconvolution. Having examined the behaviour of the value of the peak residual during the initial cleaning runs the termination threshold was set at 35,000 iterations. Following this imaging run the spectral visibility model derived from the polynomial fits to the clean component model was used to determine a set of complex gain corrections for both LL and RR polarisations (only the diagonal terms the G-Jones matrix) via self-calibration. All calibration steps were performed using the {\sc MeqTrees} package \citep{Noordam2010} using its implementation of the fast {\sc StefCal} solver \citep{Salvini2014}. 

Phase-only solutions were derived for every 300~s $\times$ 64~MHz tile of data, the frequency interval corresponding to each SPW. Solutions were forbidden from extending over the gaps in the data where the secondary calibrator observations had taken place. The calibrated data were re-imaged and the cleaning mask was refined at this stage, necessary for example if the reduced error patterns around bright sources following self-calibration revealed new genuine emission. This procedure produced acceptable images for 5 / 32 pointings. Amplitude and phase self-calibration, with additional direction-dependent calibration was required for the rest of the data. Traditional direction-independent self-calibration proved inadequate for removing the error patterns associated with off-axis sources of even modest brightness. Off-axis sources are subjected to time, frequency and direction-dependent complex gain perturbations due to the antenna primary beam response, and effect which is exacerbated by the large fractional bandwidth of the VLA, and potentially by second-order effects such as antenna pointing errors or wind loading \citep[e.g.][]{Smirnov2011b,Heywood2013a}.

Direction-dependent calibration was performed using the differential gains method \citep{Smirnov2011a}. This is an inverse-modelling approach that can be thought of as a form of simultaneous peeling \citep[e.g.][]{Noordam2004} that does not require an iterative approach, and is less prone to instabilities in the presence of confusing sources of similar brightness. Antenna-based complex gain terms (G) are derived as per traditional self-cal, based on an all-inclusive sky model, however additional solvable complex gain terms are derived for a subset of `problem' sources in the model using a longer solution interval. These additional differential gain (dE) terms are fixed to unity for all other sources.  A hybrid sky model was constructed for this purpose. Having identified the positions of the sources to which the dEs are to be applied, component-based models for these sources were derived by using {\sc PyBDSF} to characterise the emission at those positions in each of the four sub-band images produced by {\sc wsclean}. The components at these positions in the model images were then masked, and visibilities based on these model images with the problem sources removed were written to the {\tt MODEL\_DATA} column of the MS by running {\sc wsclean} in predict mode. This step makes the process computationally cheaper, as computation of the direction-\emph{independent} portion of the sky model is a one-time operation. {\sc MeqTrees} was then used to solve for G and dE terms based on the pre-computed model, plus the component models which were predicted on the fly. 

\begin{figure}
\centering
\includegraphics[width=\columnwidth]{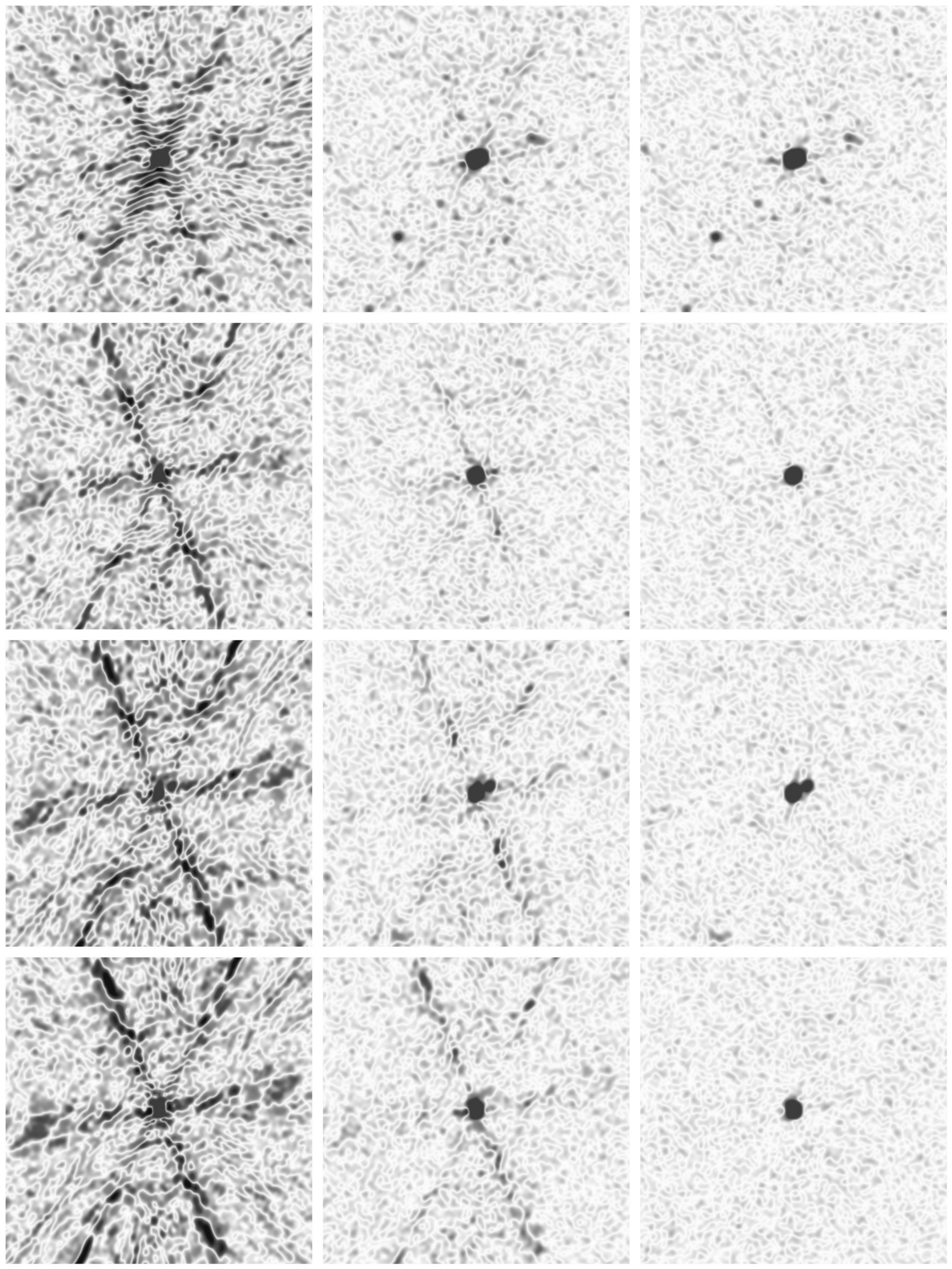}
\caption{Three generations of calibration: the results of applying different calibration schemes are shown above for four sources (one per row) selected from pointing number 1. The left hand column shows the sources as imaged following execution of the standard VLA pipeline which applies the referenced calibration. The central column shows the subsequent improvement afforded by traditional (amplitude and phase) self-calibration, and the right hand column shows the final image achieved with self-calibration with additional solvable differential gain terms applied to the four sources. These sources are ordered top to bottom by increasing radial separation from the phase centre, and all four are located somewhere between the flank of the main lobe of the primary beam and the first sidelobe, depending on the frequency. Note the degradation in the performance of self-calibration with increasing distance from the phase centre, where the primary beam related direction-dependent effects can be expected to become more pronounced. The colour scale in this image saturates black at $\pm$0.2 mJy beam$^{-1}$, with white being zero. }
\label{fig:calsteps}
\end{figure}

Solutions were derived on a per-SPW basis, with the same boundary conditions used for the phase-only solutions, and using the relevant component model for the dE terms. The solution intervals were 162 and 324~s for the G- and dE-terms respectively. These were extended by a factor of 2 for SPWs 10--12 inclusive, and by a factor of 3 for SPWs 13--15 inclusive, in order to boost the signal to noise in the solutions. Note that SPWs are zero-indexed. The cleaning masks were again refined at this stage, if required.

Figure \ref{fig:calsteps} shows the improvements in image quality gained by applying the directional calibration. The four rows correspond to four different sources in pointing 1. Top to bottom, the sources are presented in increasing distance from the phase centre. All four sources are situated between the edge of the main lobe of the primary beam and the first sidelobe, depending on the frequency. Even with the primary beam attenuation these sources are of comparable apparent brightness, and are amongst the brightest sources in the image. The first column in Figure \ref{fig:calsteps} shows the \emph{deconvolved} image produced following the application of the referenced calibration by the VLA pipeline. The second column shows the result of applying (amplitude and phase) self-calibration based on a model derived from the spectral component fitting performed by {\sc wsclean}. The third column shows the final image following the application of differential gain terms to these four sources. Note that the solution intervals for the G terms are the same for the second and third scenarios. The additional dE terms are required here to account for the differing time, frequency and direction-dependent corruptions that these sources are subjected to due to their locations in the primary beam.

Once satisfactory calibration had been performed, the data were subjected to the final imaging procedure. This made use of the final cleaning masks, with an initial constrained clean, followed by a shallower (10,000--20,000 iterations, depending on the presence of low-level extended structure) blind clean of the residual map with the mask removed. A thorough investigation of potential clean bias effects on broadband VLA snapshot data has been made by \citet{Heywood2016}. Briefly, clean (or snapshot) bias is a systematic error in the photometry measurements that is dependent on the brightness of the source being measured \citep[e.g.][]{Becker1995,Condon1997,Huynh2005}. It is thought to be related to the use of the clean algorithm for deconvolution, exacerbated by the strong linear features in the PSF of the VLA, and can even affect sources below the noise floor of the survey \citep{White2007}. The large-scale simulation conducted by \citet{Heywood2016} showed that contraining the deconvolution using masks significantly lessens the effect, but we can expect clean bias to exist at the few percent level close to the catalogue threshold, rapidly becoming negligible for brighter sources. Since a 5$\sigma$ source will be subject to statistical fluctuations at the 20\% level by definition, no corrections have been made to the catalogue for these comparatively small clean bias effects. However, persons extracting photometric measurements close to the noise floor of the survey should be mindful that clean bias may be present at the tens of percent level, comparable to the noise-induced statistical uncertainties.

A circular 2D Gaussian restoring beam with a FWHM of 4.5\arcsec was applied to each image. This is marginally broader than the generally achievable angular resolution afforded by using the fitted restoring beam, however it accounts for the variations induced in the PSF by the dynamic scheduling of the observations, and imparts a desirable uniformity to the mosaicked image. Image-plane primary beam corrections were applied to the final full-band images, as well as each of the four sub-band images, by dividing each by a model image of the VLA primary beam computed at the appropriate frequency, and masked beyond the 30\% value. Linear mosaics of the 32 images were made using the {\sc Montage}\footnote{{\tt http://montage.ipac.caltech.edu/}} package, with each pointing weighted by the assumed spatial noise variance image, in this case assumed to be represented by the square of the primary beam pattern.

\begin{figure*}
\centering
\includegraphics[width=7in]{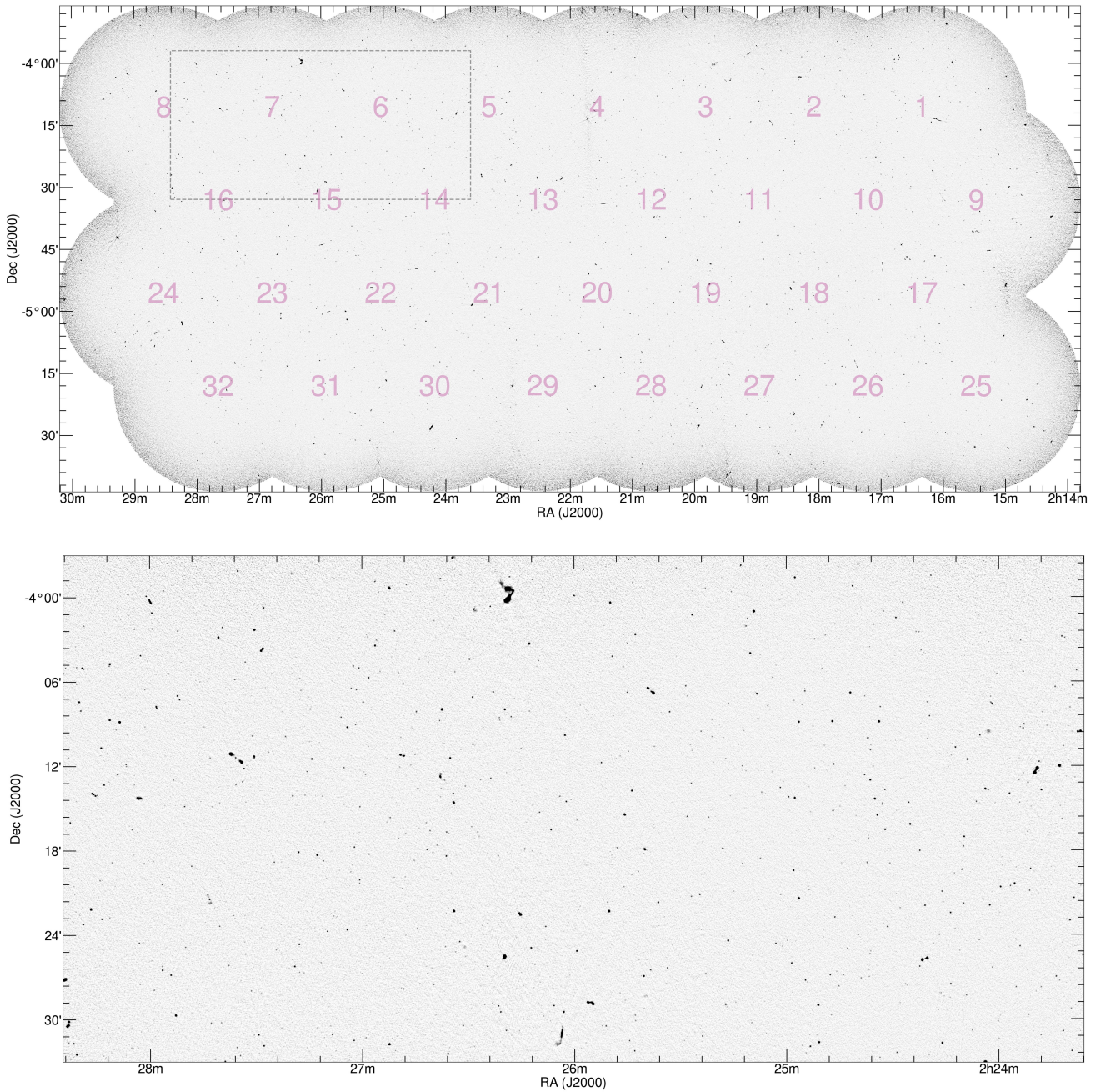}
\caption{Total intensity image formed from a linear mosaic of the 32 primary beam-corrected images (upper panel), the locations of which are indicated. The grey scale is linear and runs from $-$20 (white) to 200 $\mu$Jy~beam$^{-1}$ (black). A 1.2 $\times$ 0.6 degree region is shown in the lower panel with the same pixel scale, the location of which within the full mosaic is marked on the upper panel.}
\label{fig:mosaic}
\end{figure*}

\section{Data products}
\label{sec:data}

\subsection{Total intensity mosaic}
\label{sec:images}

The total intensity mosaic is shown in the upper panel of Figure \ref{fig:mosaic}, with each of the 32 pointing positions marked. The greyscale is linear and runs from $-$20 (white) to 20 $\mu$Jy~beam$^{-1}$ (black). The lower panel shows a 1.2 $\times$ 0.6 degree zoom on the same pixel scale, the corresponding region of which is marked by the dashed box in the upper panel. 


\subsection{Spectral index image}
\label{sec:alphaimage}

The frequency behaviour of the sky brightness distribution $I(\nu)$ is most commonly modelled as a power law in frequency
\begin{equation}
I(\nu) = I_{0}\left(\frac{\nu}{\nu_{0}}\right)^{\alpha}
\end{equation}
where $I_{0}$ is the brightness at reference frequency $\nu_{0}$, and the exponent $\alpha$ is the spectral index. For most sources over our frequency range this is a reasonable assumption. Expressing this in log-space gives
\begin{equation}
\mathrm{ln}~I(\nu)~=~\mathrm{ln}~I_{0}~+\alpha~\mathrm{ln}~\left(\frac{\nu}{\nu_{0}}\right).
\end{equation}
Defining
\begin{equation}
x~=~\mathrm{ln}~\frac{\nu}{\nu_{0}}
\end{equation}
and
\begin{equation}
y~=~\mathrm{ln}~I(\nu)
\end{equation}
allows us to compute the spectral index $\alpha$ from $N$ multi-frequency brightness measurements according to
\begin{equation}
\alpha~=~\frac{\sum_{i}(x_{i}~-~\bar{x})(y_{i}~-~\bar{y})}{\sum{i}(x_{i}~-~\bar{x})^{2}}
\end{equation}
and with standard deviation
\begin{equation}
\label{eq:alphaerror}
\sigma_{\alpha}~=~\sqrt{\frac{\sum_{i}(y_{i} - \bar{y})^{2}}{N}},
\end{equation}
where $\bar{x}$ and $\bar{y}$ are the mean values of $x$ and $y$.

Sub-band imaging of the final calibrated data is used in order to produce the multi-frequency brightness measurements required to produce an in-band spectral index image of the survey area. The 1--2 GHz band is divided up into three sections. Since SPWs 8 and 9 are discarded due to RFI in all of the 32 pointings, the LOW, MED and HIGH sub-bands are formed from SPWs 0--3, 4--7 and 10--15 inclusive. These correspond to frequency ranges of 0.994--1.25, 1.25--1.506 and 1.634--2.018 GHz, and approximately equivalent fractional bandwidths of 23\%, 19\% and 21\%. Each sub-band mosaic is formed in the same way as the full-band mosaic described in Section \ref{sec:images}, with the primary beam correction and mosaic weighting functions set by patterns appropriate to the central frequency of the sub-band. The LOW and MED mosaics are cropped to only include the high sensitivity region of HIGH, and the three images are placed into a cube with three frequency planes. The pixels in this cube are masked below 100~$\mu$Jy~beam$^{-1}$, corresponding to approximately 3--4$\sigma$ for a single sub-band, and following this a pixel-wise linear fit in log-flux/log-frequency space is performed. The best fitting gradients to each three-point spectrum are then recorded as the value of spectral index ($\alpha$) at that position, and the standard deviation in the measurements is recorded as an estimate of the spectral index error, as per Equation \ref{eq:alphaerror}. The end products of this process are maps of the spectral index and spectral index error, which we make further use of when constructing the component catalogue in Section \ref{sec:catalogues}.

\subsection{Catalogue}
\label{sec:catalogues}

\begin{table*}
\begin{minipage}{170mm}
\centering
\caption{The first ten rows from the radio source catalogue, presented here in order to show the table structure. Please refer to the text for a detailed description of each column. The full version of this table is available online as supplementary material.}\label{tab:components}
\begin{tabular}{cccccccccc}
    \hline
    &
	\multicolumn{1}{c}{ID} &
	\multicolumn{1}{c}{RA} &
	\multicolumn{1}{c}{Dec} &
	\multicolumn{1}{c}{$\sigma_{\trm{\tiny RA}}$} &
	\multicolumn{1}{c}{$\sigma_{\trm{\tiny Dec}}$} &
	\multicolumn{1}{c}{RA$_{\trm{\tiny peak}}$} &
	\multicolumn{1}{c}{Dec$_{\trm{\tiny peak}}$} &
	\multicolumn{1}{c}{$S_{\trm{\tiny int}}$} &
	\multicolumn{1}{c}{$\sigma_{S_\trm{\tiny int}}$} \\
	&
	&
	\multicolumn{1}{c}{[deg]} &
	\multicolumn{1}{c}{[deg]} &
	\multicolumn{1}{c}{[arcsec]} &
	\multicolumn{1}{c}{[arcsec]} &
	\multicolumn{1}{c}{[deg]} &
	\multicolumn{1}{c}{[deg]} &
	\multicolumn{1}{c}{[mJy]} &
	\multicolumn{1}{c}{[mJy]} \\
	&
	\multicolumn{1}{c}{(1)} &
	\multicolumn{1}{c}{(2)} &
	\multicolumn{1}{c}{(3)} &
	\multicolumn{1}{c}{(4)} &
	\multicolumn{1}{c}{(5)} &
	\multicolumn{1}{c}{(6)} &
	\multicolumn{1}{c}{(7)} &
	\multicolumn{1}{c}{(8)} &
	\multicolumn{1}{c}{(9)} \\ \hline

\textcolor{Gray}{\emph{1}}& J022143.11-041344.6 & 35.42963 & -4.22905 & 2.59 & 2.65 & 35.43002 & -4.2294 & 469.39263 & 0.01424 \\
\textcolor{Gray}{\emph{2}}& J022255.74-051817.5 & 35.73225 & -5.30485 & 2.29 & 2.07 & 35.73219 & -5.3048 & 269.69099 & 0.0188 \\
\textcolor{Gray}{\emph{3}}& J022632.54-051328.8 & 36.63557 & -5.22467 & 2.13 & 1.98 & 36.63563 & -5.22469 & 71.48008 & 0.00953 \\
\textcolor{Gray}{\emph{4}}& J022915.86-044216.7 & 37.31609 & -4.70464 & 3.94 & 3.38 & 37.31561 & -4.70498 & 272.04369 & 0.04092 \\
\textcolor{Gray}{\emph{5}}& J021640.74-044404.4 & 34.16974 & -4.73456 & 2.06 & 2.08 & 34.1698 & -4.73445 & 60.58129 & 0.00945 \\
\textcolor{Gray}{\emph{6}}& J021705.51-042253.1 & 34.27297 & -4.38143 & 1.91 & 2.29 & 34.273 & -4.38142 & 59.63392 & 0.01103 \\
\textcolor{Gray}{\emph{7}}& J022310.19-042306.4 & 35.79245 & -4.38512 & 1.95 & 1.98 & 35.79253 & -4.38508 & 39.94858 & 0.00794 \\
\textcolor{Gray}{\emph{8}}& J022754.85-045705.5 & 36.97856 & -4.95152 & 2.24 & 2.0 & 36.9785 & -4.95146 & 35.40523 & 0.00853 \\
\textcolor{Gray}{\emph{9}}& J022357.09-044112.5 & 35.98789 & -4.68682 & 1.99 & 2.35 & 35.98794 & -4.68674 & 42.21077 & 0.00876 \\
\textcolor{Gray}{\emph{10}}& J022505.11-053648.1 & 36.27128 & -5.61335 & 2.2 & 2.44 & 36.27124 & -5.61345 & 161.54529 & 0.03284 \\

 \hline
\end{tabular}
\vspace{2mm}
\begin{tabular}{rcccccccccccc} \hline
	&
	\multicolumn{1}{c}{$S_{\trm{\tiny peak}}$} &
	\multicolumn{1}{c}{$\sigma_{S_\trm{\tiny peak}}$} &
	\multicolumn{1}{c}{$\trm{RMS\_Peak}$} &
	\multicolumn{1}{c}{$\trm{RMS\_Mean}$} &
	\multicolumn{1}{c}{$\theta_{\mathrm{maj}}$} &
	\multicolumn{1}{c}{$\theta_{\mathrm{min}}$} &
	\multicolumn{1}{c}{PA} &
	\multicolumn{1}{c}{$\alpha$} & 
	\multicolumn{1}{c}{$\sigma_{\alpha}$} &
	\multicolumn{1}{c}{ID2} &
	\multicolumn{1}{c}{ID3} \\

&
	\multicolumn{1}{c}{[mJy~b$^{-1}$]} &
	\multicolumn{1}{c}{[mJy~b$^{-1}$]} &
	\multicolumn{1}{c}{[mJy~b$^{-1}$]} &
	\multicolumn{1}{c}{[mJy~b$^{-1}$]} &
	\multicolumn{1}{c}{[arcsec]} & 
	\multicolumn{1}{c}{[arcsec]} &
	\multicolumn{1}{c}{[deg]} &
	\multicolumn{1}{c}{} &
	\multicolumn{1}{c}{} &
	\multicolumn{1}{c}{} &
	\multicolumn{1}{c}{} \\
	\\
 &
	\multicolumn{1}{c}{(10)} &
	\multicolumn{1}{c}{(11)} &
	\multicolumn{1}{c}{(12)} &
	\multicolumn{1}{c}{(13)} &
	\multicolumn{1}{c}{(14)} &
	\multicolumn{1}{c}{(15)} &
	\multicolumn{1}{c}{(16)} &
	\multicolumn{1}{c}{(17)} &
	\multicolumn{1}{c}{(18)} &
	\multicolumn{1}{c}{(19)} &
	\multicolumn{1}{c}{(20)} & \\ \hline

\textcolor{Gray}{\emph{1}}& 301.11002 & 0.00914 & 0.02774 & 0.02761 & 24.06 & 14.99 & 136.56   & -0.62& 0.16   & --&--  \\
\textcolor{Gray}{\emph{2}}& 232.57525 & 0.01621 & 0.02984 & 0.02973 & 23.57 & 21.31 & 95.65   & -0.27 & 0.03   & --&--  \\
\textcolor{Gray}{\emph{3}}& 68.15745 & 0.00909 & 0.0185 & 0.01847 & 20.27 & 17.55 & 60.05   & -0.65   & 0.21   & --&--  \\
\textcolor{Gray}{\emph{4}}& 153.31504 & 0.02306 & 0.05453 & 0.05453 & 32.7 & 18.35 & 53.51   & --      & --   & --&--  \\
\textcolor{Gray}{\emph{5}}& 52.00014 & 0.00811 & 0.02091 & 0.02088 & 17.02 & 16.45 & 34.36   & -0.68  & 0.27   & --&--  \\
\textcolor{Gray}{\emph{6}}& 55.30472 & 0.01023 & 0.02289 & 0.02282 & 19.34 & 16.16 & 0.34   & -0.69   & 0.15   & --&--  \\
\textcolor{Gray}{\emph{7}}& 37.44547 & 0.00744 & 0.01913 & 0.01915 & 15.77 & 14.88 & 38.87   & -0.32  & 0.17   & --&--  \\
\textcolor{Gray}{\emph{8}}& 32.4521 & 0.00782 & 0.01786 & 0.01786 & 18.84 & 16.49 & 105.35   & -0.42  & 0.10   & --&--  \\
\textcolor{Gray}{\emph{9}}& 33.05087 & 0.00686 & 0.01939 & 0.01936 & 18.47 & 15.06 & 17.55   & -1.18  & 0.23   & --&--  \\
\textcolor{Gray}{\emph{10}}& 116.68966 & 0.02372 & 0.06663 & 0.06639 & 19.22 & 17.14 & 167.76  & --   & --   & --&--  \\
\hline
\end{tabular}
\end{minipage}
\end{table*}

The package \textsc{ProFound} \citep{Robotham2018} was used to generate an associated source catalogue from the total intensity mosaic. Although designed for optical/near-IR surveys, \textsc{ProFound} has been shown to be able to successfully model radio emission \citep{Hale2019b} for sources of different morphologies. As \textsc{ProFound} does attempt to fit to any particular morphology (e.g 2D Gaussians), complex morphologies (e.g. AGN with extended jets) may be more faithfully modelled. 

To extract the source catalogue, the method of \cite{Hale2019b} is followed. We use a \texttt{skycut} value of 3.5, which only includes pixels that have a value of $3.5 \times$ the sky RMS value at that pixel within a source segment. The segment defines all the pixels of a source that contribute to the model for the source. As the source density does not approach that of the classical confusion limit, the \texttt{groupstats=TRUE} setting is used to force neighbouring segments that share a segment boundary to be combined into a single source. This is especially important for resolved extended sources that for example have connected lobe emission, and ensures that (provided the emission is connected) these sources can be identified as a single source. \\

Following the method of \cite{Hale2019b} we apply a (restoring) beam correction to ensure that emission within the wings of the source (especially for faint point-like sources) is not missed. To do this, we take all segments below a given pixel threshold limit and investigate what fraction of the total flux contained within the PSF beam and centred on the RA/Dec position of the source is contained within the source segment. We apply this correction to those sources which have a value of \textsc{N100} (the number of pixels in the segment found by \textsc{ProFound}) less than 225 pixels. This limit is chosen as 225 pixels in a 15 x 15 pixel box around a central PSF should contain $\sim 99\%$ of the total flux within a PSF beam.

Using \textsc{ProFound} with these settings resulted in a catalogue of 7,185 sources. We subsequently discard fitted regions where the peak flux density is below five times the noise value at the peak position of the source, resulting in a final catalogue of 5,780 sources. After a visually examining images, we identified 13 sources for which multiple components (a total of 30) were actually a single association. For these sources, the associated components are recorded within the final table. In addition to these, 18 sources were deemed to be artefacts and were subsequently removed from the catalogue. Following this, a total of 5,762 sources remained within the final catalogue. The properties of the first ten sources from our final catalogue are shown in Table \ref{tab:components}. The columns are defined as follows:\\

\noindent
(1) Identifier for the component in HHMMSS.S+/-DDMMSS format, formed from the right ascension and declination position in the J2000 epoch.\\
\noindent
(2-3) Flux-weighted right ascension and declination of the component in degrees taken from the \texttt{RAcen} and \texttt{Deccen} columns from \textsc{ProFound}.\\
\noindent
(4-5) Flux-weighted standard deviations in the right ascension and declination of the component, taken from the \texttt{xsd} and \texttt{ysd} columns from \textsc{ProFound} and converted into angular units using the pixel sizes. Note that this is significantly larger than the statistical uncertainty that can be obtained by fitting a point or Gaussian component, and is included here mainly for completeness.\\
\noindent
(6-7) Right ascension and declination of the peak of the source in degrees taken from the \texttt{RAmax} and \texttt{Decmax} columns from \textsc{ProFound}.\\
\noindent
(8) Integrated flux density of the component in mJy. This is calculated using the \textsc{ProFound} \texttt{flux} column, converted to Jy (from Jy beam$^{-1}$), with an appropriate beam correction applied to compensate for the flux density contribution from the outer wings of the emission (see text). \\
\noindent
(9) Error in the integrated flux density of the component in mJy. It is calculated similar to (3) but using \texttt{flux\_err} instead of \texttt{flux} and applying the square root of the beam correction.\\
\noindent
(10) Peak intensity of the component in mJy beam$^{-1}$. This is constructed from the \textsc{ProFound} catalogue as \texttt{flux}$\times$\texttt{cenfrac}.\\
\noindent
(11) Error in the peak intensity of the component in mJy beam$^{-1}$. It is calculated similar to (7) but using \texttt{flux\_err} instead of \texttt{flux}.\\
\noindent
(12) RMS value in the map at the peak position of the source (given by columns 6-7).\\
(13) Mean rms over the source segment using the \texttt{skyRMS\_mean} column from \textsc{ProFound}.\\
\noindent  
\noindent  
(14) Major axis size of the segment and is quoted here as the 2$\times$\texttt{R100} column from \textsc{ProFound} and converted to arcseconds.\footnote{As this (and the minor axis size) are calculated based on the segment size, for faint sources comparable to the noise, the segment will be small and this size will be underestimated. These are also not comparable (in many cases) to sizes in previous radio catalogues, which are often quoted as full width half maximum values from Gaussian components.}\\
\noindent  
(15) Minor axis size of the segment and is quoted here as the 2$\times$\texttt{R100}$\times$\texttt{axrat} from \textsc{ProFound} and converted to arcseconds.\\
\noindent  
(16) Positional angle of the source in degrees given by the \texttt{ang} column from \textsc{ProFound}.\\
\noindent  
(17) Spectral index ($\alpha$) estimate formed by extracting pixels from the spectral index map (Section \ref{sec:alphaimage}) over the region corresponding to a given source as determined by \textsc{ProFound}. The mean of the spectral index value of the extracted pixels is determined, weighted by the total intensity values over the same area.\\
\noindent  
(18) Total intensity weighted standard deviation of $\alpha$, measured over the corresponding \textsc{ProFound} region.\\
(19)-(20) IAU Source IDs of components that together with the entry in column (1) are part of a single radio source.\\

\section{Results and discussion}
\label{sec:discussion}

In the sections that follow we compare the results presented in Sections \ref{sec:images} and \ref{sec:catalogues} with existing radio data in order to validate these data products. For positional and flux density checks (Sections \ref{sec:astrometry} and \ref{sec:photometry}) we make use of existing data covering the same field, namely the Faint Images of the Radio Sky at Twenty-cm (FIRST) survey \citep{Becker1995}, and radio imaging of the VLA-VIRMOS Deep Survey field \citep[VVDS;][]{Bondi2003}, and Subaru-XMM/Newton Deep Field \citep[SXDF;][]{Simpson2006}. For astrometric and photometric checks (Sections \ref{sec:astrometry} and \ref{sec:photometry}) we restrict the cross-match to sources that have no clear evidence of having extended morphology in order to minimise the effects of angular resolution differences.

\subsection{Sensitivity}
\label{sec:sensitivity}

\begin{figure}
\centering
\includegraphics[width= \columnwidth]{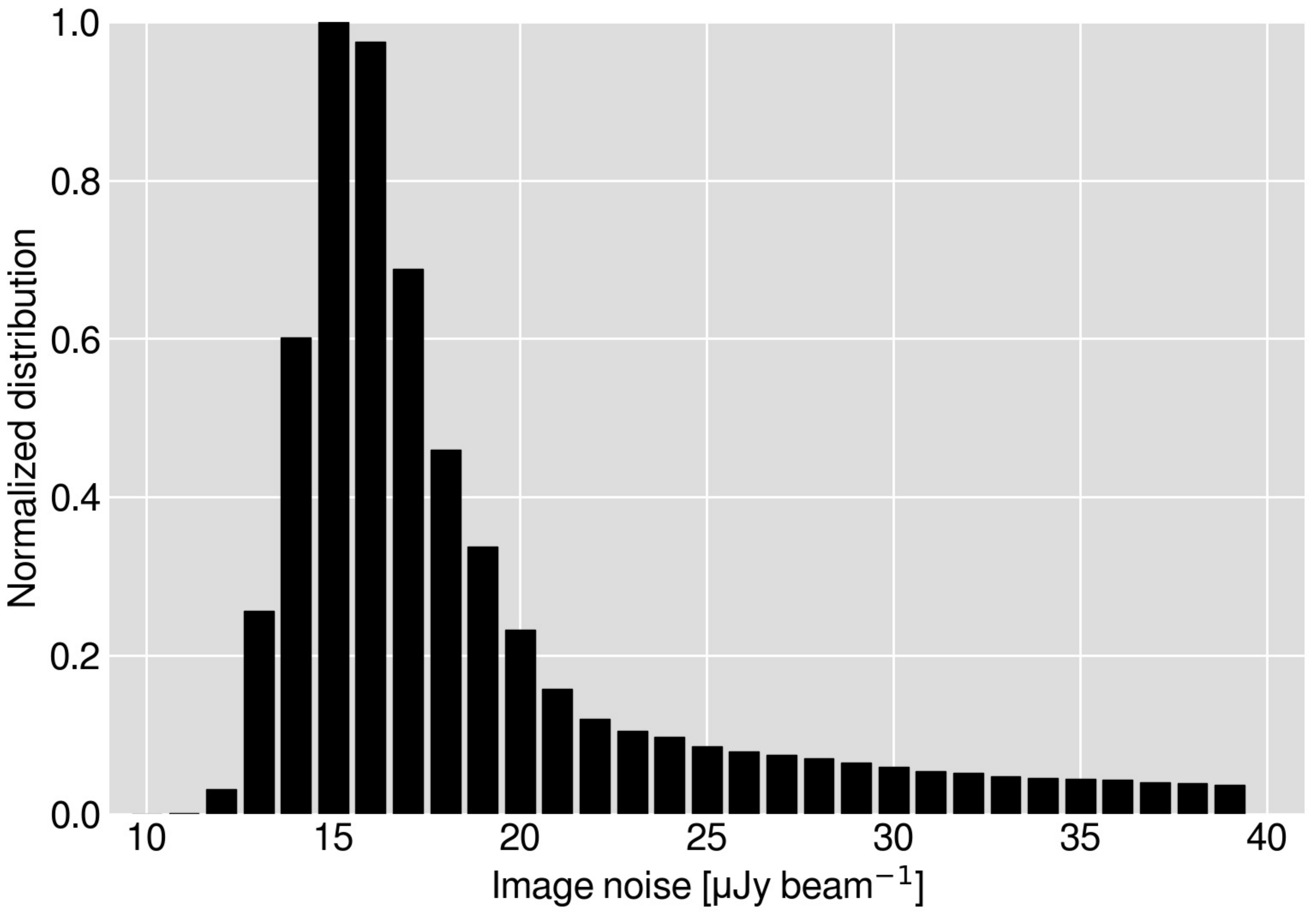}
\caption{Normalised histogram of the pixel values in the RMS image of the mosaic, taken to be a measurement of the background noise across the survey. The median noise is 16~$\mathrm{\mu}$Jy~beam$^{-1}$.}
\label{fig:rmshisto}
\end{figure}

The sensitivity (or background noise level) of a radio mosaic at these frequencies is generally position dependent. This can be due to a range of factors, e.g.~the increase of the noise at the periphery of the mosaic due to primary beam correction, calibration deficiencies leading to error patterns associated within bright sources (e.g. Figure \ref{fig:calsteps}), residual sidelobe confusion due to incomplete deconvolution, and particularly problematic RFI in some pointings causing higher than normal data loss for that region. A convenient way to capture the sensitivity of the mosaic as a function of position is to make use of the RMS noise map that is produced by the source finder in order to set its internal local detection thresholds. Figure \ref{fig:rmshisto} shows the a normalised histogram of the pixels in this RMS image. The median RMS noise is 16~$\mathrm{\mu}$Jy~beam$^{-1}$, with 80\% of the mosaic area having a noise value of $<$20~$\mu$Jy~beam$^{-1}$.

\subsection{Astrometry}
\label{sec:astrometry}

\begin{figure*}
\centering
\includegraphics[width=7in]{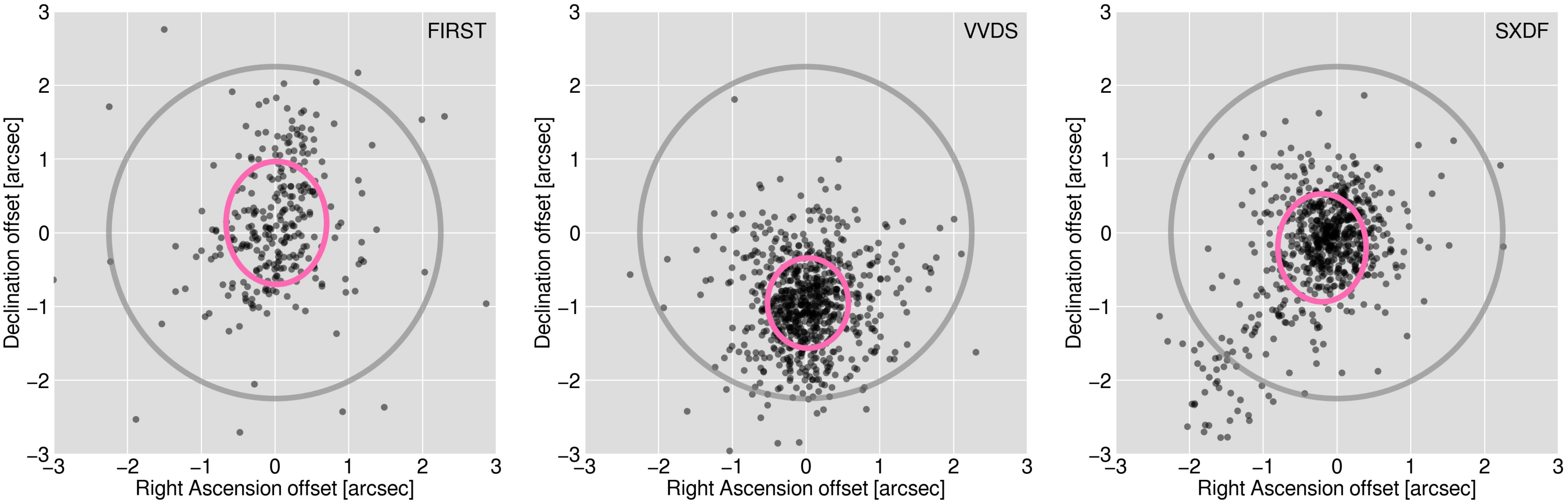}
\caption{Differences in the \emph{peak} right ascension and declination of components matched between the {\sc ProFound} catalogue derived in Section \ref{sec:catalogues} and an external reference data set. Left to right, the external references are FIRST \citep{Becker1995}, VVDS \citep{Bondi2003} and SXDF \citep{Simpson2006}. The inner ellipse is centred on the mean positions of the offsets, and its major and minor axes are $\pm$1 standard deviation of the offsets in right ascension and declination. The outer circle shows the extent of the restoring beam used in our final mosaic. Systematic offsets are $\sim$ 1 arcsecond or better in all cases. The mean $\pm$ 1 standard deviation offsets in right ascension and declination with respect to the reference observations are as follows: FIRST (333 sources), RA offset 0.052 $\pm$ 0.561 arcseconds, Dec offset 0.128 $\pm$ 0.793 arcseconds; VVDS (724 sources), RA offset -0.004 $\pm$ 0.549 arcseconds, Dec offset -0.944 $\pm$ 0.662 arcseconds; SXDF (690 sources), RA offset -0.170 $\pm$ 0.613 arcseconds, Dec offset -0.150 $\pm$ 0.729 arcseconds.}
\label{fig:positions}
\end{figure*}

The accuracy to which the position of a component in a radio image can be measured depends on two factors \citep{Condon1997}. The first is a statistical effect related to the signal to the noise ratio (SNR) of the detection and the angular resolution of the instrument. The second is a systematic component coupled to accuracy of the astrometric reference frame that is applied to the data via the calibration. Both of these effects can be gauged by cross-matching positional measurements with those of suitable reference data, if such data are available. Ideally the reference set should have superior depth and angular resolution such that the statistical uncertainties are dominated by those of the survey under test. The systematic calibration-related component is best investigated by using the strongest sources (e.g.~phase calibrators) for which the statistical contribution in both data sets is negligible. In practice, and with many modern radio observations breaking new ground, the availability of suitable reference sets is limited, and typically relies on using a large-area survey to investigate the brighter sources that are common to both. The use of bright calibrator sources with excellent positional measurements is generally not feasible for deep and relatively narrow surveys such as the one presented here, however the astrometry of surveys such as FIRST and NVSS is validated against calibrator sources, so with a large enough sample of common sources any systematic offsets should be apparent.

We calculate offsets in right ascension and declination between the \emph{peak} positions in our catalogue and the matched position in an external reference catalogue. Three external catalogues are employed, namely FIRST, VVDS and SXDS. The distribution of these offsets is shown in Figure \ref{fig:positions}. The inner ellipse is centred on the mean positional offset, and has minor and major axes showing $\pm$1 standard deviation in the distribution in right ascension and declination. The mean position and standard deviations are noted in the caption of Figure \ref{fig:positions}, along with the number of matched components.The outer circle shows the FWHM of the 2D Gaussian restoring beam used during imaging. In each case the mean offsets are less than 1 arcsecond, corresponding to less than 25\% of the FWHM of the effective angular resolution of the final mosaic.

The tail of sources in the lower left of the SXDF panel on Figure \ref{fig:positions} was investigated further, and $\sim$90\% of them were found to lie within the bounds of pointing 7 of the SXDF mosaic, suggesting an issue either with the calibration or image regridding for that particular pointing. The offsets between our catalogue positions and those of VVDS is noticeable compared to those of FIRST and SXDF, however given that we are consistent with the latter two we assume this is related to the VVDS calibration.

\subsection{Photometry}
\label{sec:photometry}

\begin{figure}
\centering
\includegraphics[width= \columnwidth]{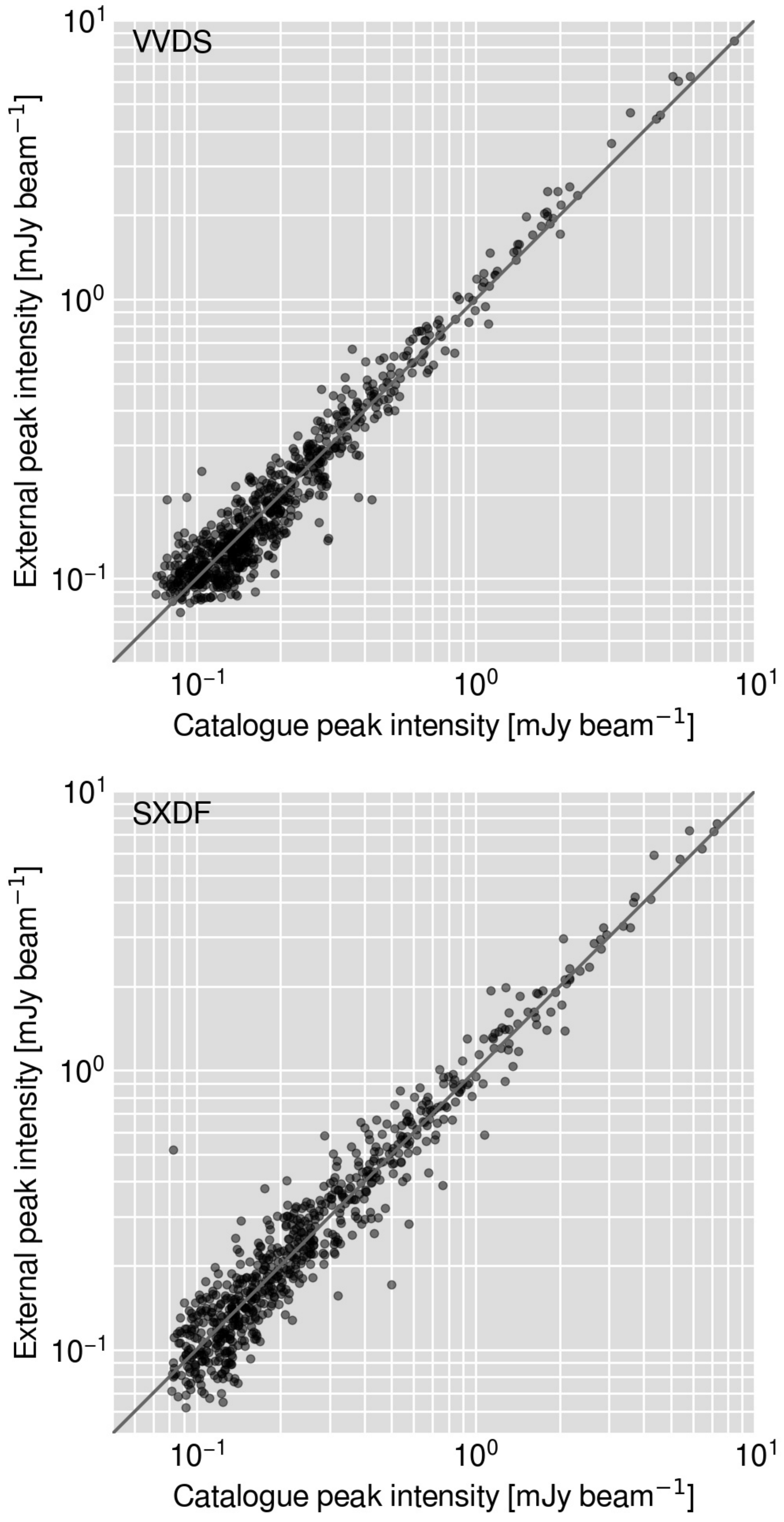}
\caption{Comparison of the peak intensities from two external radio surveys, namely VVDS \citep{Bondi2003} and SXDF \citep{Simpson2006}, plotted against the catalogued peak intensity from our survey.}
\label{fig:photometry}
\end{figure}

The VLA has very accurate absolute flux calibration \citep[of order 1 percent;][]{Perley2013} due to the use of well-modelled primary calibrator sources, in this case 3C147. However, additional factors (e.g.~subsequent referenced calibration and self-calibration problems, deconvolution biases, RFI) can skew the flux calibration. In Figure \ref{fig:photometry} we compare the peak flux densities of our catalogues components with matched components drawn from the SXDS and VVDS catalogues. As with the positional checks there were 690 and 724 mutually-compact sources for SXDF and VVDS respectively. 

Matched components are scattered about the 1:1 line where the catalogued and external component are equal, as shown as the diagonal on Figure \ref{fig:photometry}. The usual increase in scatter with decreasing peak intensity is seen. As the noise level becomes an increasingly large fraction of the component brightness temporally separate measurements of the same source will exhibit larger amounts of scatter. There is no obvious biasing of e.g.~the fainter sources, as would be seen by a curve in the distribution of points about the 1:1 line.

A potential source of bias in the recovered flux density of sources is the application of inappropriate self-calibration. Since the sky model against which the instrument is calibrated is never fully complete, the contribution to the visibility function made by the unmodelled sources can potentially be absorbed by the antenna-based gain solutions, resulting in these unmodelled sources being suppressed in the final image. Mitigation of this effect can take the form of conservative time-frequency solution intervals, and minimising the degrees of freedom. The latter issue is automatically addressed to some extent by virtue of the VLA having a high ratio of baselines (351) to antennas (27), which results in a correspondingly high ratio of equations to solvable parameters during calibration. However the application of differential gains introduces two solvable parameters into the measurement equation for every additional direction that is being solved for, and extra care must be taken.

\begin{figure}
\centering
\includegraphics[width= \columnwidth]{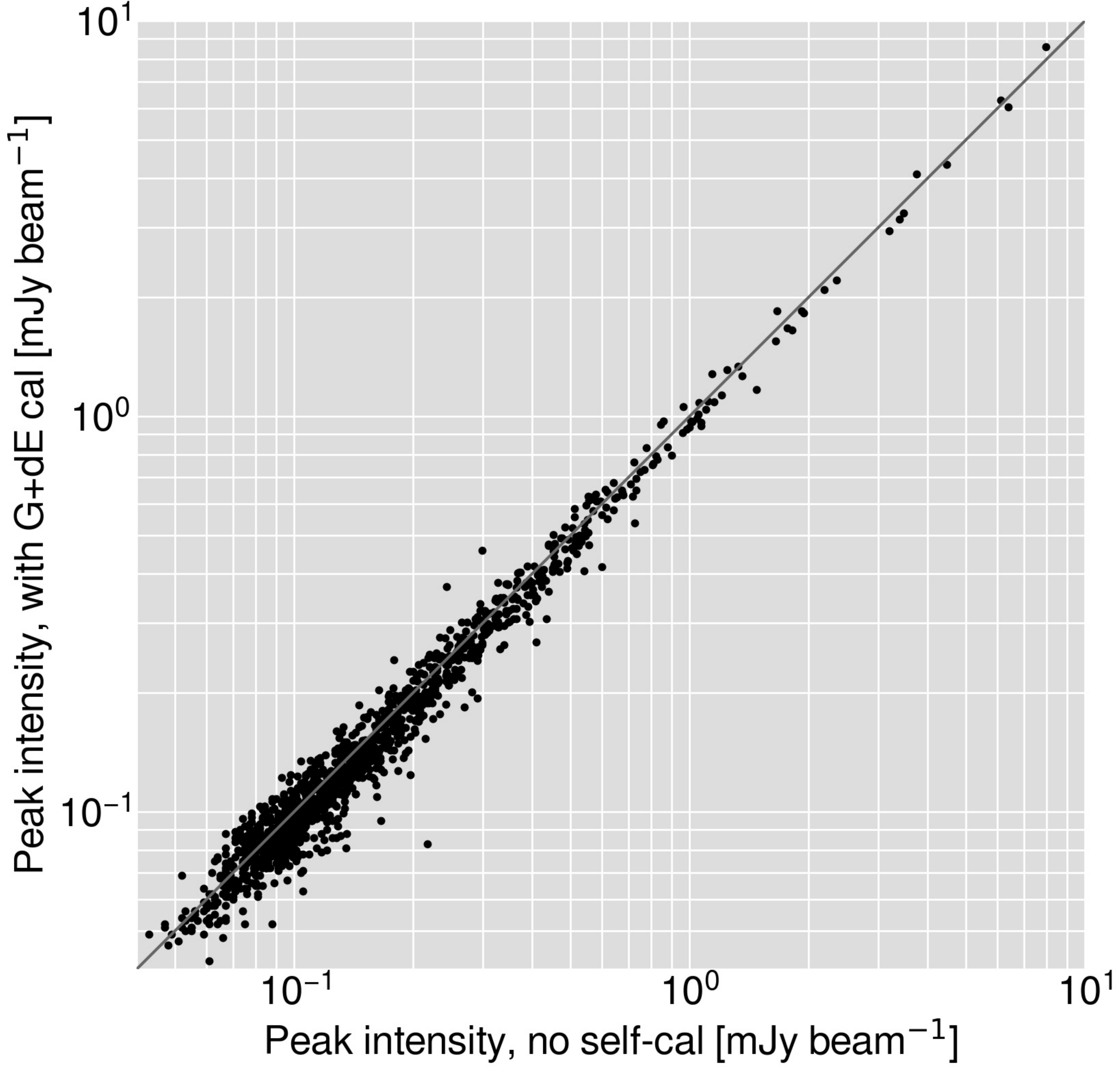}
\caption{Peak intensities from the final calibrated maps of some individual pointings plotted against the peak intensities of matched components measured from maps that have not been subjected to any self-calibration, with only the referenced calibration applied. There is no evidence for any calibration-induced photometry biases. Note that the diagonal line here is a fit to the data points.}
\label{fig:suppression}
\end{figure}

We check for the presence of systematic flux density biases introduced by the directional calibration process by comparing the flux densities of matched components in the images formed from the NRAO pipeline (i.e. maps for which no self-calibration has been applied) and those formed following the full direction-dependent calibration procedure. These are plotted in Figure \ref{fig:suppression}. 

As with Figure \ref{fig:photometry}, this plot shows increased broadening of the distribution away from the diagonal line with decreasing values of component flux density. Note however that in this case the diagonal line is not simply the 1:1 line, but rather a fit to the data that is indistinguishable from the diagonal. A noise-like scattering of the points is to be expected, as self-calibration modifies the noise properties of the images. Systematic biasing of the flux density measurements (for example the often-seen suppression of faint sources that are not in the calibration model) would manifest itself as a curve in the distribution, for which no evidence is seen.

\subsection{Spectral indices}
\label{sec:spectroscopy}

\begin{figure}
\centering
\includegraphics[width= \columnwidth]{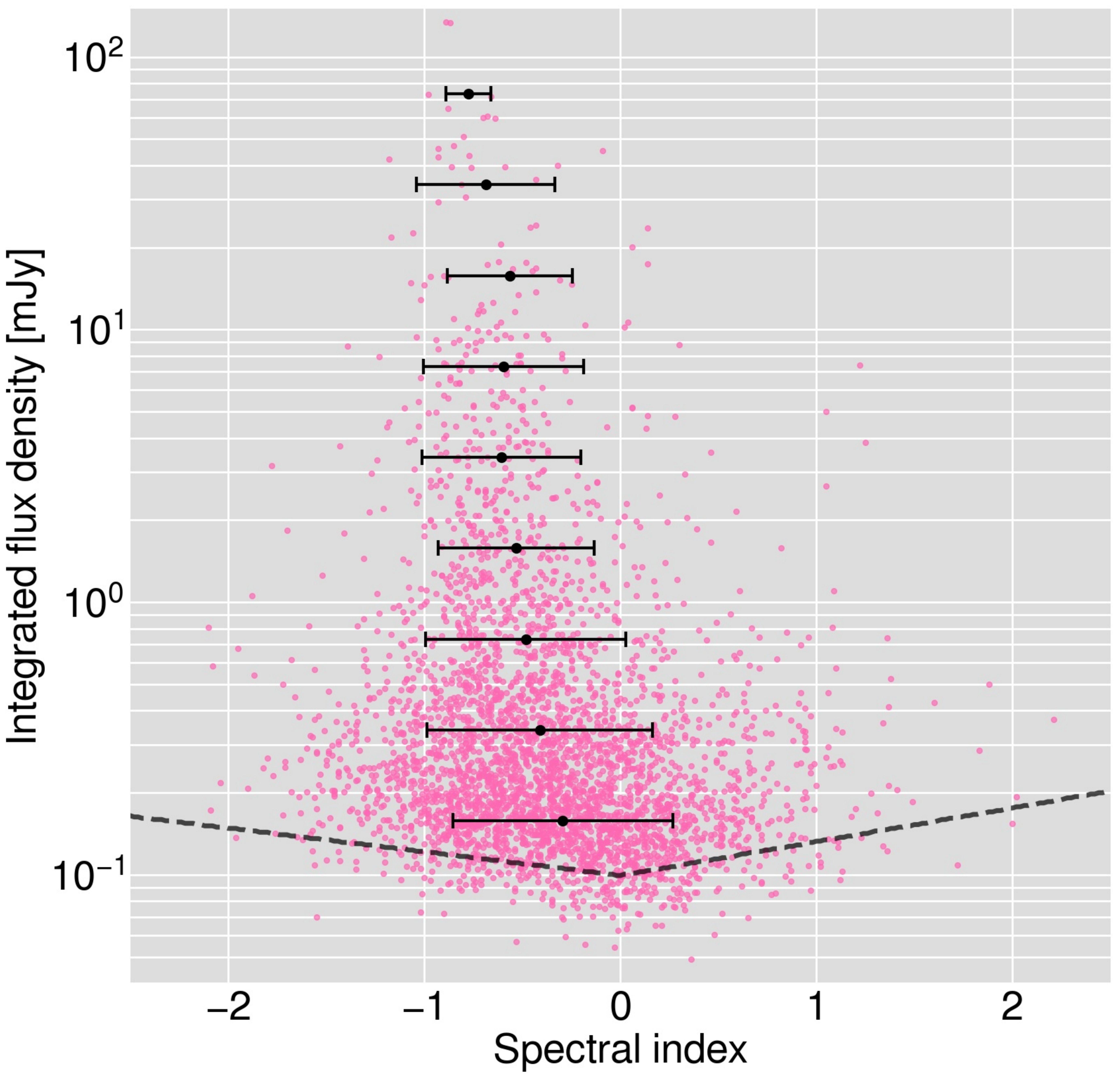}
\caption{Integrated flux density measurement against source spectral index for the 3,458 sources that have spectral index estimates. The black markers show the mean spectral index for the flux density bins listed in Table \ref{tab:medalphas}, with the error bars showing 1 standard deviation. The dashed line shows the theoretical \emph{peak} flux density limit \emph{(as would be measured in the full-band image)} below which a source of a given spectral index would drop out of one of the three sub-bands used to form the spectral index map. Sources above the entirety of this line can be assumed to be mostly free from signal-to-noise related selection biases for most common or plausible astrophysical radio spectra.}
\label{fig:alphas}
\end{figure}

The catalogued spectral index measurements (having their origins described in Sections \ref{sec:alphaimage} and \ref{sec:catalogues}) are shown in Figure \ref{fig:alphas}, with the corresponding integrated flux density plotted against them. The dashed line on this plot shows the limit where a source with a peak intensity in mJy beam$^{-1}$, concordant with the y-axis values and measured from the full-band mosaic, would drop below the threshold of any one of the three sub-bands used, and therefore not have a spectral index measurement in this survey. This line is evaluated for the full spectral index range of Figure \ref{fig:alphas}, and demonstrates that in-band spectral indices are subject to a spectral index dependent selection bias for sources with flux densities that approach the survey detection threshold. The population of sources that lie above this line can thus be considered to be complete for plausible and typical spectral indices.

The mean spectral index of the sources in integrated flux density bins is measured, and these values are plotted on Figure \ref{fig:alphas} with error bars that show $\pm$1 standard deviation. The mean spectral index values per bin are listed in Table \ref{tab:medalphas}. A tendency towards flatter mean spectral index measurements with decreasing flux density is seen, with the mean value changing from -0.6 to -0.4 over the order of magnitude drop in integrated flux density between 3.4 and 0.34 mJy (see also Table \ref{tab:medalphas}). This flattening trend is consistent with previously reported in-band measurements \cite[e.g.][]{Heywood2016} and dual-frequency measurements between 1.4 and 5~GHz \cite[e.g.][]{Prandoni06}, however we note that such a trend has not been seen in dual-frequency spectral index studies between 1.4~GHz and lower frequencies, e.g. the 610~MHz work of \citet{Ibar09}. \citet{Huynh2015} present a 5~GHz mosaic covering 0.34 deg$^{2}$ of the Extended Chandra Deep Field South to a depth of 8.6 $\mu$Jy beam$^{-1}$, from which they match 167 sources with counterparts from 1.4 GHz VLA observations reaching 6 $\mu$Jy beam$^{-1}$ \citep{Miller2013}. A flattening of the median spectral index (-0.58) is observed in the sub-mJy population, however a substantial fraction of flat or inverted spectrum radio sources is present.

\begin{table}
\centering
\caption{Mean spectral index values of the N components with integrated flux densities within the range defined by $S_{min}$ (inclusive) and $S_{max}$ (with bin centre $S$). These values are plotted as black markers on Figure \ref{fig:alphas}. Further details are provided in Section \ref{sec:spectroscopy}.}
\begin{tabular}{cccccc} \hline
N & $S_{min}$  & $S_{max}$  & $S$ & $\alpha_{\mathrm{med}}$ & $\sigma_{\alpha}$\\ 
    &$[$mJy b$^{-1}]$ & $[$mJy b$^{-1}]$ & $[$mJy b$^{-1}]$ &   & \\ \hline
1386 & 0.1    & 0.215  & 0.158  & -0.293 & 0.562 \\
984  & 0.215  & 0.464  & 0.34   & -0.411 & 0.575 \\
433  & 0.464  & 1.0    & 0.732  & -0.483 & 0.511 \\
217  & 1.0    & 2.154  & 1.577  & -0.532 & 0.399 \\
125  & 2.154  & 4.642  & 3.398  & -0.606 & 0.405 \\
80   & 4.642  & 10.0   & 7.321  & -0.596 & 0.409 \\
32   & 10.0   & 21.544 & 15.772 & -0.563 & 0.318 \\
18   & 21.544 & 46.416 & 33.98  & -0.687 & 0.352 \\
8    & 46.416 & 100.0  & 73.208 & -0.774 & 0.115 \\ \hline
\end{tabular}
\label{tab:medalphas}
\end{table}

\subsection{Differential source counts and bias corrections}

Next, we measure the differential source counts for these observations and compare them to previous work. Before comparisons can be made with previous studies it is important to correct the measured differential source counts, which will be underestimated especially at the faintest flux densities. This underestimation is due to several factors. Firstly the variations in the image sensitivity across the survey area means that faint sources will not be detectable in all regions of the image. Secondly, false detections whereby noise peaks are interpreted as true emission will affect the source counts in the faintest bins. Furthermore, sources for which any (positive or negative) coincident noise peak represents an appreciable fraction of their total flux density may be redistributed into an adjacent bin (Eddington bias). The methods for determining the factors required to correct for these effects are described below.\footnote{The catalogues used to determine these corrections also have the 5$\sigma$ criterion described in Section \ref{sec:catalogues} imposed before the corrections is calculated.}

\subsubsection{Completeness correction}
\label{sec:sc_comp}

Firstly, we correct the measured source counts from the output catalogue for non-uniform detection across the field of view as well as results that may arise from source fluxes being influenced by noise peaks or troughs. To determine the necessary corrections, simulations are used to correct source counts as in \cite{Hale2019a} and the corrections determined are applied to this work. For this, simulated sources are injected into the image and then source extraction is run as in Section \ref{sec:catalogues}, this can be used to to determine the recovery as a function of flux density. For each simulation, 1000 sources were injected at random positions within the image. Each of these sources has an associated flux density with the distribution drawn from the SKA Simulated Skies continuum simulation \citep[$S^3$][]{Wilman2008, Wilman2010}, which provides realistic catalogues containing simulated extragalactic radio sources of various population types down to a flux limit of 10 nJy. The shapes of the injected sources are elliptical components, with the associated sizes also drawn from the $S^3$ simulations. For each simulated source randomly chosen, each elliptical component associated with the source are convolved with clean beam of these observations and then injected into the image\footnote{$S^3$ sources which had components of the largest sizes were not included, to ensure the lobes are not cut off when injecting into the image. Sources with sizes>50" were not included. Only a small fraction of $S^3$ sources were not included due to this limit and so are unlikely to have made a big difference to the corrections derived.}. As this uses the size distribution of $S^3$, this technique should also account for resolution effects where, for the same total flux density, larger sources will have a lower peak flux density per beam and therefore will be more challenging to detect above the noise threshold. Both single (radio quiet AGN and star-forming galaxies) and multi-component (FR-I and FR-II radio galaxies) sources are injected into the image provided that the total flux of the source is $\geq 3 \times \sigma$, where $\sigma$ was taken as the typical rms of the observations (converted to a total flux assuming a point source), and was taken to be 16 $\mu$Jy beam$^{-1}$.

To determine the completeness from the simulation we calculate the ratio of the output source count distribution to the input distribution. As the simulated sources are injected into the image, the observed (real) source count distribution measured from the image must first be subtracted. The completeness correction in a given flux density bin is therefore given by:

\begin{equation}
    C_{\mathrm{COMP}}(S_i, S_i +dS_i) = \frac{N_{\mathrm{sim, out}}(S_i, S_i +dS_i) - N_{\mathrm{im}}(S_i, S_i +dS_i)}{N_{\mathrm{sim, in}}(S_i, S_i +dS_i)}
    \label{eq:comp}
\end{equation}

\noindent where $N_{\mathrm{sim,out}}$ is the number of sources detected in the output simulated image above 5$\sigma$ (as defined in Section \ref{sec:catalogues}, $N_{\mathrm{im}}$ is the number of sources within the original image, again above 5$\sigma$  and finally $N_{\mathrm{sim, in}}$ is the number of simulated sources within the given flux density bin that are injected into the image. As $N_{\mathrm{sim,out}}$ will be the combination of both the sources already in the image as well as those simulated sources that are recovered, the value of $N_{\mathrm{sim,out}}~-~N_{\mathrm{im}}$ will quantify those simulated sources that are recovered from the image. As sources that are injected into the image, these simulations may also take into account the fact that sources may merge with others in the image and only be detectable as an individual source. 

We generated 100 realisations of the simulation and calculated the completeness corrections as the median value of these. The associated uncertainties that we quote with this are generated from the 16$^{\textrm{th}}$ and 84$^{\textrm{th}}$ percentiles of the completeness corrections for the 100 simulations. The inverse of these corrections will need to be applied to the measured source counts in order to correct for incompleteness.

\subsubsection{False detection correction}
\label{sec:sc_fdr}
To quantify the fraction of false detections in the image, we make the assumption that the noise across the image is symmetric and therefore every positive noise spike will on average have a corresponding noise decrement. As such, the number of falsely detected sources within a given flux density bin can be calculated through investigating how many sources would be detected within the negative image (i.e. where the image is multiplied by -1). The same detection parameters of \textsc{ProFound} (as described in Section \ref{sec:catalogues}), 5$\sigma$ threshold and beam correction method \citep[as described in][]{Hale2019b} are used to extract the catalogue of sources in the negative image. As this correction aims to account for the fact that some sources within the measured catalogue may be false, this correction will act to decrease the measured source counts. This is in the opposite direction to the corrections described in Section \ref{sec:sc_comp}. The correction that is applied to account for these false detections is given by:

\begin{equation}
    C_{\mathrm{FDR}} (S_i, S_i +dS_i)= 1 - \frac{N_{\mathrm{neg}}(S_i, S_i +dS_i)}{N_{\mathrm{cat}}(S_i, S_i +dS_i)}
\end{equation}

\noindent where $N_{\mathrm{cat}}$ is the number of sources within the flux density bin $S_i, S_i +dS_i$ and $N_{\mathrm{neg}}$ is the number of sources within the same flux density bin that are detected within the negative image.

\begin{table*}
\begin{minipage}{170mm}
\centering
\caption{The data corresponding to Figure \ref{fig:sourcecounts} for measurements below 500 mJy. For each flux density bin we list the differential source counts in both raw and Euclidean-normalised form. The final column lists the source counts following the application of the corrections described in Section \ref{sec:sc_comp} and \ref{sec:sc_fdr}.}\label{tab:sourcecounts}
\begin{tabular}{cccccccccc}
\hline
	\multicolumn{1}{c}{Bin} &
	\multicolumn{1}{c}{Bin Mid} &
	\multicolumn{1}{c}{Counts} &
	\multicolumn{1}{c}{Raw $\frac{dN}{dS}S^{2.5}$} & 
	\multicolumn{1}{c}{Corrected $\frac{dN}{dS}S^{2.5}$} & \\
	\multicolumn{1}{c}{[mJy]} &
	\multicolumn{1}{c}{[mJy]} &
	\multicolumn{1}{c}{} &
	\multicolumn{1}{c}{[sr$^{-1}$ Jy$^{1.5}$]} &
	\multicolumn{1}{c}{[sr$^{-1}$ Jy$^{1.5}$]} &\\ \hline 

0.10 - 0.13 & 0.11 & 636 $\pm$ 25 & 1.45 $\pm$ 0.06 & $3.94 ^{+0.78}_{-0.67} $\\ 
 0.13 - 0.16 & 0.14 & 688 $\pm$ 26 & 2.22 $\pm$ 0.08 & $4.27 ^{+1.09}_{-0.83} $\\ 
 0.16 - 0.20 & 0.18 & 604 $\pm$ 24 & 2.75 $\pm$ 0.11 & $4.85 ^{+0.96}_{-0.85} $\\ 
 0.20 - 0.25 & 0.22 & 450 $\pm$ 21 & 2.89 $\pm$ 0.14 & $4.19 ^{+0.70}_{-0.85} $\\ 
 0.25 - 0.32 & 0.28 & 368 $\pm$ 19 & 3.34 $\pm$ 0.17 & $4.44 ^{+0.94}_{-0.71} $\\ 
 0.32 - 0.40 & 0.35 & 340 $\pm$ 18 & 4.36 $\pm$ 0.23 & $5.49 ^{+1.13}_{-0.83} $\\ 
 0.40 - 0.50 & 0.45 & 240 $\pm$ 15 & 4.35 $\pm$ 0.27 & $5.00 ^{+1.16}_{-0.89} $\\ 
 0.50 - 0.63 & 0.56 & 178 $\pm$ 13 & 4.56 $\pm$ 0.33 & $5.14 ^{+1.58}_{-0.92} $\\ 
 0.63 - 0.79 & 0.71 & 159 $\pm$ 12 & 5.75 $\pm$ 0.43 & $6.13 ^{+1.74}_{-0.95} $\\ 
 0.79 - 1.00 & 0.89 & 118 $\pm$ 10 & 6.03 $\pm$ 0.51 & $6.91 ^{+1.58}_{-1.70} $\\ 
 1.00 - 1.26 & 1.12 & 102 $\pm$ 10 & 7.36 $\pm$ 0.72 & $8.17 ^{+2.18}_{-1.62} $\\ 
 1.26 - 1.58 & 1.41 & 78 $\pm$ 8 & 7.95 $\pm$ 0.82 & $7.75 ^{+2.14}_{-1.45} $\\ 
 1.58 - 2.00 & 1.78 & 67 $\pm$ 8 & 9.65 $\pm$ 1.15 & $9.65 ^{+2.67}_{-1.68} $\\ 
 2.00 - 2.51 & 2.24 & 58 $\pm$ 7 & 11.79 $\pm$ 1.42 & $11.79 ^{+2.98}_{-2.43} $\\ 
 2.51 - 3.16 & 2.82 & 60 $\pm$ 7 & 17.24 $\pm$ 2.01 & $17.24 ^{+6.09}_{-3.20} $\\ 
 3.16 - 3.98 & 3.55 & 44 $\pm$ 6 & 17.85 $\pm$ 2.43 & $17.85 ^{+5.29}_{-3.14} $\\ 
 3.98 - 5.01 & 4.47 & 36 $\pm$ 6 & 20.63 $\pm$ 3.44 & $20.63 ^{+4.86}_{-5.37} $\\ 
 5.01 - 6.31 & 5.62 & 34 $\pm$ 5 & 27.53 $\pm$ 4.05 & $27.53 ^{+10.03}_{-5.36} $\\ 
 6.31 - 7.94 & 7.08 & 34 $\pm$ 5 & 38.88 $\pm$ 5.72 & $38.88 ^{+9.65}_{-8.80} $\\ 
 7.94 - 10.00 & 8.91 & 31 $\pm$ 5 & 50.08 $\pm$ 8.08 & $50.08 ^{+18.54}_{-14.90} $\\ 
 10.00 - 12.59 & 11.22 & 21 $\pm$ 4 & 47.92 $\pm$ 9.13 & $47.92 ^{+19.52}_{-9.13} $\\ 
 12.59 - 15.85 & 14.13 & 13 $\pm$ 3 & 41.90 $\pm$ 9.67 & $41.90 ^{+9.67}_{-9.67} $\\ 
 15.85 - 19.95 & 17.78 & 12 $\pm$ 3 & 54.63 $\pm$ 13.66 & $54.63 ^{+13.66}_{-13.66} $\\ 
 19.95 - 25.12 & 22.39 & 5 $\pm$ 2 & 32.15 $\pm$ 12.86 & $32.15 ^{+12.86}_{-12.86} $\\ 
 25.12 - 31.62 & 28.18 & 5 $\pm$ 2 & 45.42 $\pm$ 18.17 & $45.42 ^{+18.17}_{-18.17} $\\ 
 31.62 - 39.81 & 35.48 & 4 $\pm$ 2 & 51.32 $\pm$ 25.66 & $51.32 ^{+25.66}_{-25.66} $\\ 
 39.81 - 50.12 & 44.67 & 11 $\pm$ 3 & 199.37 $\pm$ 54.37 & $199.37 ^{+54.37}_{-54.37} $\\ 
 50.12 - 63.10 & 56.23 & 6 $\pm$ 2 & 153.61 $\pm$ 51.20 & $153.61 ^{+51.20}_{-51.20} $\\ 
 63.10 - 79.43 & 70.79 & 4 $\pm$ 2 & 144.65 $\pm$ 72.33 & $144.65 ^{+72.33}_{-72.33} $\\ 
 100.00 - 125.89 & 112.20 & 1 $\pm$ 1 & 72.15 $\pm$ 72.15 & $72.15 ^{+72.15}_{-72.15} $\\ 
 125.89 - 158.49 & 141.25 & 1 $\pm$ 1 & 101.92 $\pm$ 101.92 & $101.92 ^{+101.92}_{-101.92} $\\ 
 158.49 - 199.53 & 177.83 & 1 $\pm$ 1 & 143.97 $\pm$ 143.97 & $143.97 ^{+143.97}_{-143.97} $\\ 
 251.19 - 316.23 & 281.84 & 2 $\pm$ 1 & 574.51 $\pm$ 287.25 & $574.51 ^{+287.25}_{-287.25} $\\ 
 316.23 - 398.11 & 354.81 & 1 $\pm$ 1 & 405.76 $\pm$ 405.76 & $405.76 ^{+405.76}_{-405.76} $\\ 
 398.11 - 501.19 & 446.68 & 1 $\pm$ 1 & 573.15 $\pm$ 573.15 & $573.15 ^{+573.15}_{-573.15} $\\ \hline

\end{tabular}
\end{minipage}
\end{table*}

\subsubsection{Corrected Source Counts}
To obtain the corrected source counts, which should be a true estimate of the underlying flux density distribution, the corrections from the completeness simulations and false detections are combined together multiplicatively to the source counts determined from the measured output catalogue described in Section \ref{sec:catalogues}. This is applied to the source count from the catalogue where artefacts have not been removed. This is because these artefacts may also be apparent in the inverted image for which the false detection rate is determined from. The associated uncertainties from the measured source counts and the corrections are then combined together in quadrature in order to quantify the total uncertainty on these corrected source counts. The corresponding measurements of the source counts and their uncertainties are given in Table \ref{tab:sourcecounts}. 

\begin{figure*}
\centering
\includegraphics[width= 0.8\textwidth]{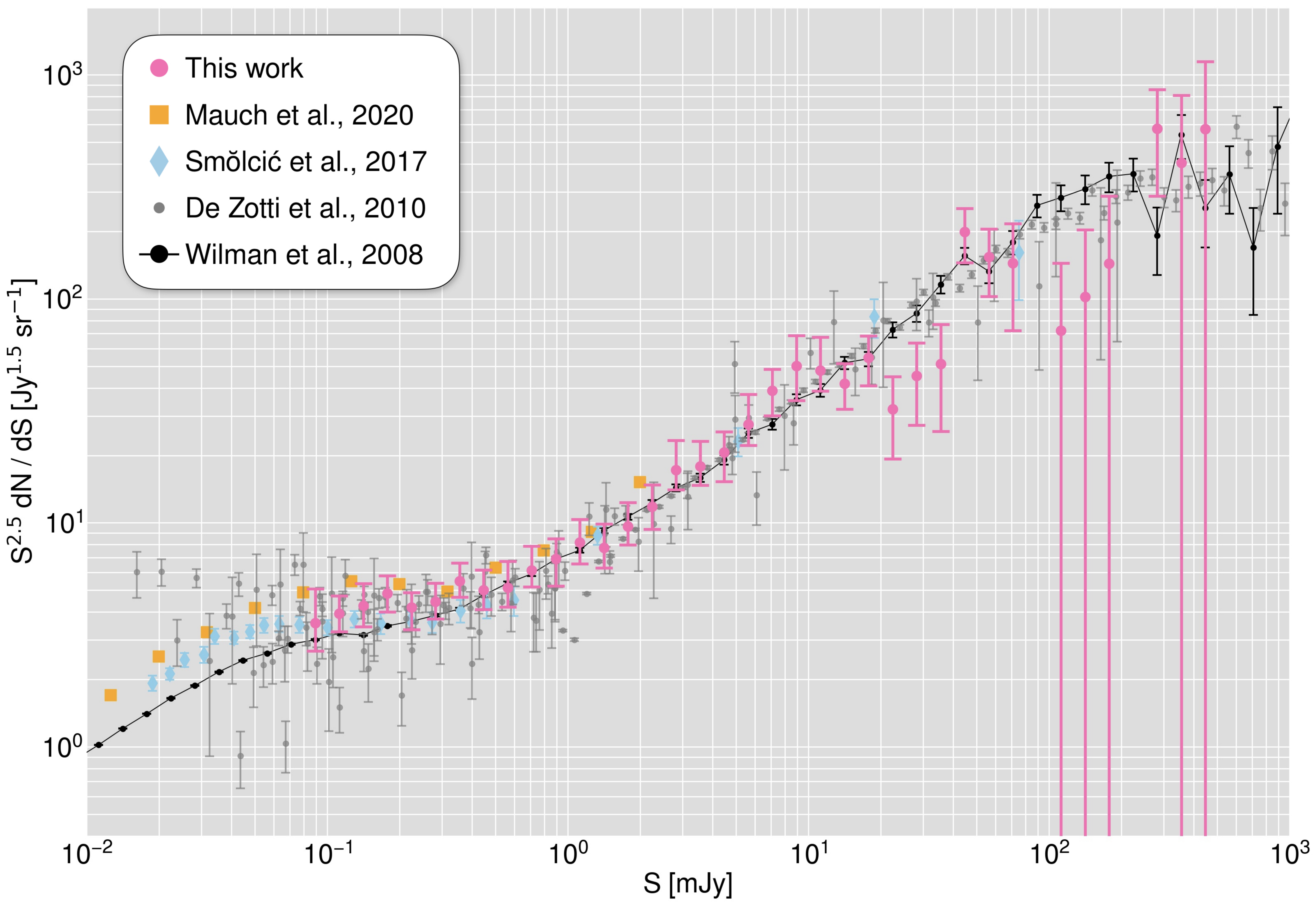}
\caption{Euclidean normalised differential source counts and their associated uncertainties for these observations are shown above via the pink markers circles. Also shown are the comparisons to previous work from simulations at 1.4 GHz of \protect \cite{Wilman2008} (black markers) and also the compilation of observations from \protect \cite{deZotti2010} (grey markers), the VLA 3 GHz COSMOS Survey from \protect \cite{Smolcic2017} (blue markers), and the 1.28 GHz MeerKAT observations from \protect \citet{Mauch2020}. These are all scaled to 1.4 GHz assuming a spectral index of -0.7 where necessary.}
\label{fig:sourcecounts}
\end{figure*}

A comparison of the uncorrected and corrected source counts are presented in Figure \ref{fig:sourcecounts}, for which observations of previous measured source counts are also presented. These previous source count measurements are from the VLA 3 GHz COSMOS Survey \citep{Smolcic2017}, the compilation of 1.4 GHz source counts presented by \cite{deZotti2010} and finally the $S^3$ extragalactic simulated skies at 1.4 GHz \citep{Wilman2008}. These are all scaled to 1.4 GHz assuming a spectral index of -0.7. The source counts are plotted for those flux density bins that have a minimum value greater than the 5$\sigma$ limit (where $\sigma$ is taken as 16 $\mu$Jy). As can be seen from Table \ref{tab:sourcecounts}, the corrections to the source counts become important at flux densities of $S \lesssim 0.7$ mJy. At these lower flux densities, the corrections applied appear to successfully correct the measured source counts plotted in Figure \ref{fig:sourcecounts}, in between those of previous observations from \cite{Smolcic2017} and \citet{Mauch2020}, and the simulations of \cite{Wilman2008}. The observations reach the regime where the flattening or upturn in the measured source counts is seen, thought to represent the emergence of the star forming galaxy population seen in the radio via their optically-thin synchrotron emission \citep[e.g.][]{Condon2012}.
At higher flux densities, the measured source counts are also comparable with previous work, although we note the dip in the counts at around 30 mJy. Perturbations to the source count measurements due to field-to-field variations (sample variance) are expected to be negligible for the faint end of a survey of this area and depth \citep{Heywood2013b}. However, as readily be seen from Figure \ref{fig:sourcecounts}, similar effects can skew the counts at the bright end and this is likely the simplest explanation for the 30 mJy dip.

\section{Conclusion}
\label{sec:conclusion}

We have described the production and validation of the reduced data products associated with a VLA survey covering $\sim$5 deg$^{2}$ of the XMM-LSS / VIDEO field. The data we present enhance the multi-wavelength view of one of the best-studied extragalactic deep fields, and we make these products publicly available for use by the community, downloadable from http://tiny.cc/vla-xmm, or by emailing the contact author.

Direction-dependent calibration has been used to produce a broadband radio mosaic that reaches a thermal noise-limited median depth of 16 $\mathrm{\mu}$Jy beam$^{-1}$ with an angular resolution of 4.5$''$. Our survey improves on the existing matched-frequency radio data, expanding the area by a factor of 2.5 to encompass the entire region for which the deep near-infrared VIDEO data \citep{Jarvis2013} are available, and further increase the depth of the radio data available over this region at these frequencies by 40\%.

A source catalogue with 5,762 entries has been produced using the {\sc ProFound} source finder, recently demonstrated to have excellent performance for the characterisation of extended radio sources by \citet{Hale2019b}. The photometric and astrometric performance of the resulting catalogue (and thus the radio mosaic from which it is derived) have been validated by comparison to existing narrowband observations. The bias-corrected differential source counts are also in excellent agreement with simulations and observations. The 66\% fractional bandwidth of the VLA allows in-band spectral indices to be estimated for sources detected at sufficiently high signal to noise ratios, and the catalogue contains spectral index measurements for 60\% of sources. The mean spectral index as a function of integrated flux density resembles the canonical synchrotron values at the bright end, tending towards flatter spectrum sources below about 1~mJy.

Looking forwards, a second data release will be forthcoming with observations from the compact C and D configurations of the VLA being added in order to improve the sensitivity to the many diffuse structures that are evident in the mosaic. Observations using new mid-frequency SKA precursor instruments will also target this field, with the MIGHTEE survey \citep{MIGHTEE} specifically planning deep observations of XMM-LSS to greater depths at comparable frequencies. The superior angular resolution of our VLA data will prove useful not only for validation of the MIGHTEE data, but also potentially for disentangling confused sources for the optical cross-identification at 100 $\mathrm{\mu}$Jy beam$^{-1}$ and above.


\section*{Acknowledgements}
\addcontentsline{toc}{section}{Acknowledgements}

We thank the anonymous referee and the MNRAS editorial staff for their comments. The National Radio Astronomy Observatory is a facility of the National Science Foundation operated under cooperative agreement by Associated Universities, Inc. We thank the VLA Director for awarding us 1.5~h of Director's Discretionary Time in order to complete this project. This work was supported by resources provided by the Pawsey Supercomputing Centre with funding from the Australian Government and the Government of Western Australia. IH thanks the Rhodes Centre for Radio Astronomy Techniques and Technologies (RATT), South Africa, for the provision of computing facilities. This research has made use of NASA's Astrophysics Data System. This research made use of Montage. It is funded by the National Science Foundation under Grant Number ACI-1440620, and was previously funded by the National Aeronautics and Space Administration's Earth Science Technology Office, Computation Technologies Project, under Cooperative Agreement Number NCC5-626 between NASA and the California Institute of Technology. This research made use of APLpy, an open-source plotting package for Python \citep{aplpy}. This work was supported by the Oxford Hintze Centre for Astrophysical Surveys which is funded through generous support from the Hintze Family Charitable Foundation. CLH acknowledges the Science Technology and Facilities Council (STFC) for their support through an STFC Studentship.




\bibliographystyle{mnras}
\bibliography{vid} 





\bsp	
\label{lastpage}
\end{document}